\documentclass[12pt,preprint]{aastex}

\begin{document}


\title{Millisecond Pulsar Searches and Double Neutron Star Binaries}

\author{J. Middleditch}
\affil{Modeling. Algorithms, and Informatics, MS B265, Computer \& Computational \\
       Science Division, Los Alamos National Laboratory, Los Alamos, NM 87545}

\email{jon@lanl.gov}

\begin{abstract}
A unified strategy is developed that can be used to 
search for millisecond pulsars (MSPs) with $\sim$solar mass companions 
(including neutron star companions in double neutron
star binaries [DNSBs]) belonging to both very short period
binaries, and those with periods so long that they could
be appropriate targets for acceleration searches, and to
bridge the gap between these two extremes.
In all cases, the orbits are assumed to be circular.
Applications to searches for binary pulsars similar
to PSR J0737-3039 are discussed.  The most likely
candidates for more DNSBs consist of
weakly magnetized neutron stars, spinning only
moderately fast, like J0737-3039A, with periods
generally longer than 15 ms, though this issue
is not yet settled.  Because of the similarity between the 
MSP components of DNSBs, and the longer period MSP population
specific to massive condensed or core collapsed globular 
clusters, as well as the uncertainties about accretion-driven
spinup, doubts linger about the standard models of 
DNSB formation.

\end{abstract}


\section{Introduction}

The discovery of the double binary pulsar system PSR J0737-3039A,B  
(hereafter J0737 -- Burgay et al.~2003; Lyne et al.~2004) has
rekindled interest in computation intensive searches for other 
similar systems, because their coalescence events will be 
detectable as sources to gravitational observatories \citep{Kal04}, 
that will most likely be stronger than merger-induced core collapse 
supernovae (SNe) if very much less common (Middleditch 2004a, 
hereafter M04a).  None of the other seven double neutron star binaries
(DNSBs) have detected pulsar companions.  Four of these
(B1534+12, J1756-2251, B1913+16, and M15C) have 
orbital periods of 10, and 3*8 hours respectively 
\citep{W91,Fau04,HT75,Pri91} one has a period of 28 hours 
(J1829+2456, Champion et al.~2004) and the remaining two 
(J1518+4904 and J1811-1736) have periods of 9 and 19 days
(Nice, Sayer, and Taylor 1996; Lyne et al.~2000).  In addition, three 
(J1811-1736, B1913+16, and M15C) have eccentricities of 0.83, 0.62, and 
0.68, while four (J1518+4904, B1534+12, J1756-2251, and J1829+2456) have 
eccentricities of 0.25, 0.27, 0.18, and 0.14.
By contrast, the eccentricity for J0737, with 
an orbital period of only 2.4 hours, is only 0.088.  
Should this trend continue as DNSBs evolve, then 
deeper, brute force searches using only circular orbits
and quantum computers if necessary, may ultimately 
find  more (however, the lifetime of such close binary systems
is very short).  Other applications include millisecond pulsar 
(MSP) searches in low mass X-ray binaries (e.g., Dib et al.~2004), 
and very young supernova (SN) remnants (SNRs) in which a newly-born pulsar is 
spinning down rapidly and/or mimicking binary motion through 
precession (Nelson, Finn, \& Wasserman 1990; Middleditch et al.~2000,
hereafter M00).  

Even membership in a binary system with circular 
orbits, however, can drastically reduce the detectability of an
MSP to conventional search techniques.  In fact, if
J0737A\&B were moderately weaker at 1,400 MHz than
1.6 and 1.3 mJy, they would never have been detected even 
by acceleration searches.  In addition, 
through the use of radio imaging techniques at sites such 
as the VLA, and others in Australia, point sources will be 
mapped that will be pulsar candidates, but will also be near the 
threshold of detection as pulsars even by conventional 
acceleration searches.  Thus, many such mapped sources will also
require brute force binary pulsar searches to uncover their
pulsar parameters.

The next section ($\S 2$) begins with a discussion of acceleration
searches ($\S 2.1$), continues with a discussion of short
period binary searches ($\S 2.2$), and derives how to transition
between the two search extremes.  Section 2.3 gives an example
of a short period binary search, and $\S 2.4$
discusses ways to maintain the effectiveness of such searches 
with much less required computation.  The following sections
($\S 3$ and $\S 4$) discuss how the DNSBs could affect MSP 
searches, and conclude.

\section{Technique}

\subsection{Acceleration Searches}

The problem to be solved by future searches, then, is how to
compensate, {\it efficiently, but still completely}, for the possibility
that a target being searched may be a pulsar member of a short period binary
system, and how to extend such searches into the range of
longer binary periods.  By ``short'' we mean a binary period that
is short enough so that the drift in pulse frequency due to the
pulsar's acceleration in the binary system can be greater than one
or two wavenumbers of frequency $\rm{(cycles~(observation~time)^{-1})}$.
Expressed in terms of the physical parameters
of the binary system, this constraint becomes:
\begin{equation}
{f'_{max}T^{2}=1.844*10^{-5}(f/650)T^{2}P_{d}^{-4/3}\sin i
(M/2.4M_{\odot})^{1/3}2.4M_{c}/M\geq1}~,
\end{equation}
where $f$ is the pulsar repetition frequency in Hz,
$f'_{max}$ is the maximum absolute value of
the time derivative of the pulse frequency caused by
the orbital motion, $T$ is the time duration of the
observation of the pulsar, $P_{d}$ is the orbital
period in days, sin$i$ is the sin of the
orbital inclination, $i$, and $M$ is the total mass of the
binary system, i.e., the mass of the pulsar, $M_{p}$,
plus the mass of the companion star, $M_{c}$.
Equation (1) shows that even an observation as short as 
one(five) minute(s) for a 0.1/1.0 day
binary period, can be
affected for a 650 Hz pulsar orbiting a solar
mass companion.

Thus we need to know how densely we must
sample the parameters of the binary orbit in
order that we lose, in the worst case situation,
only 10\%, 20\%, or 40\% of the
power that a pulsar would produce in a Fourier
transform of a given observation, relative to
the power produced {\it if} the orbital parameters 
of our data-taking apparatus (or of our time
series-stretching software) were {\it perfectly}
matched to the orbital parameters of the pulsar.
In order to solve this problem, it is first
easier to investigate the problem in the
limit of the ``longer'' short orbital periods,
where we need only concern ourselves with
matching the first two time derivatives of
frequency, $f'$ and $f''$.

The errors that can be associated
with the measurement of the rate of frequency
drift in time, $f'$, and the next time derivative,
$f''$, are given by:
\begin{equation}
\delta f' = \sqrt{90}/(\pi \sqrt{{P_F}}T^{2})~,
\end{equation}
\begin{equation}
\delta f'' = 6 \sqrt{105}/(\pi \sqrt{{P_F}}T^{3})~,
\end{equation}
where $P_F$ is the Fourier power produced by the periodic
pulsations in a power spectrum normalized
so that the local mean power near (but not at) the
pulsar frequency (or harmonic being discussed) is
unity.  These errors apply only when $f'$ and $f''$ are to
be measured while the remaining parameters, $f$ and $f''$, or
$f$ and $f'$ respectively are known a priori to the measurement.
The derivation of the
formulae for $\delta f$
(which is given by $\delta f = 3 /(\pi \sqrt{6{P_F}}T )$)
and $\delta f'$ are outlined elsewhere (Ransom, Eikenberry, 
\& Middleditch 2002, hereafter REM02);
the formula for $\delta f''$ can be derived with similar
arguments.  (The formula for $\delta f'''$ can not be so derived,
possibly because pure $f'''$ is almost indistinguishable
from an ordinary periodic frequency modulation, which produces
little effect at the central frequency.  However the 2nd harmonic of
the 2.14 ms pulsations from SN1987A during 31 Oct.~`95 was so 
highly phase modulated that the Fourier peak at the central frequency 
was starting to {\it split} --  M00.)  Using a continuous Fourier interpolation 
and the first, second, and third derivatives of the Fourier amplitudes
wrt the frequency ($f$), $f'$, and $f''$ can be estimated for a small
range about 0, from the Fourier amplitudes close to $f$ 
(Middleditch \& Cordova 1982; Middleditch, Deich, \& Kulkarni 1992, 
hereafter MDK).

Since a power spectrum with unit mean has the statistical properties of
one half of the $\chi^{2}$ for two degrees of freedom, we could have
defined the errors given in equations (2) and (3) as the deviations
of the parameters $f'$ and $f''$ (taken one at a time) which reduced
the power of a Fourier peak by 0.5 units below the maximum value.
Thus we can rework equations (2) and (3) to solve for the deviations
which reduce the peak power by 10\% (i.e., reduced to 0.9 of the peak
value):
\begin{equation}
T^{2} \delta f'_{90\%} = 3 \sqrt{2}/\pi~,
\end{equation}
\begin{equation}
T^{3} \delta f''_{90\%} = 6 \sqrt{21}/\pi~.
\end{equation}
By differentiation of the Doppler equation for $f$,
we can relate $f'$ and $f''$ to a function of the orbital
phase, $\Omega t$:
\begin{equation}
f' = { f \Omega ^{2} a_p \sin i \over c} \cos (\Omega t)~,
\end{equation}
\begin{equation}
f'' = { f \Omega ^{3} a_p \sin i \over c} \sin (\Omega t)~,
\end{equation}
where $t$ is time, $c$ is the velocity of light,
$\Omega = 2 \pi / P$, where $P$ is the orbital period,
and $a_{p}$ is the size of the
orbit of the pulsar in the binary system:
\begin{equation}
{a_{p} = 1.633 \ast 10^{11} P_{d}^{2/3} ({M \over 2.4M_{\odot}})^{1/3}
\ast{2.4 M_{c}/M}~~ (\rm cm)~.}
\end{equation}
We can see from equations (6) and (7) that $f'$ and $f''$ represent
orthogonal coordinates of the orbital system (ignoring sin $i$ for the
moment), with $f'$ measuring the depth into the binary system
along the projected line of sight \citep{MN76},
and $f''$
measuring the extent of the binary system in the plane of the
sky, perpendicular to the conjunction axis (i.e., along the
line of quadratures).

Thus, if we wanted to search our data for a
pulsar in a not-too-short orbital period system,
we would superpose a rectangular lattice of points
over a circular orbit, where the lattice spacing
is given by twice the tolerances of equations (4)
and (5).
At quadrature, with both binary members at maximum elongation
(and thus in the plane of the sky,
or $\Omega t = {\pi \over 2}~or~{3 \pi \over 2}$),
$f'$ is near 0, but changing very rapidly, and hence $f''$ 
is near one of its two extremes, but changing only very
slowly.  Here also
the {\it differentials} of $f'$ and $f''$ resolve into
relations that depend only upon the differentials of
one of the two binary parameters, $a_{p}$ and
orbital phase, $\Phi = \Omega t$:
\begin{equation}
{{\vert \delta f' \vert}_{max} = {\delta \Phi f \Omega^{2} a_{p} \sin i \over 2c}~;
~(\Phi = {\pi \over 2}, {3\pi \over 2})},
\end{equation}
\begin{equation}
{\vert \delta f'' \vert}_{max} = {\delta a_{p} \sin i f \Omega^{3}  \over 2c};~~~~~~~~"~~~~~.
\end{equation}
Using equations (4) and (5) for the tolerances, we
can derive an expression for the tolerances of the
binary parameters, $a_{p}$, and $\Phi$:
\begin{equation}
\delta \Phi = {6 \sqrt{2} c \over \pi f (\Omega T)^{2} a_{p} \sin i}~~~~(radians)~,
\end{equation}
\begin{equation}
\delta a_{p} = {12 \sqrt{21} c \over \pi f (\Omega T)^{3} \sin i}
;~~(\Phi = {\pi \over 2}, {3\pi \over 2})~.
\end{equation}
The above equations give the mismatch of the orbital parameters
that would produce a 10\% loss of the power of a Fourier
peak (per parameter).
 
The simplicity of equations (11) and (12)
has provided part of the
motivation behind the quadrature search technique.
This involves scanning
through a data set large enough to contain at least one quadrature, plus
at least half of the analysis size length, $T/2$, and tracking the data
through the orbital motion, with increments in orbital parameters
close to those given by equations (11) and (12), hoping to detect
the pulsar.
The other part of the motivation for the quadrature
search has been due, in part, to computing limitations,
either because of an upper limit to the size of the
Fourier transforms that could be done, or
an upper limit to the total amount of computing time
available.
Here, however, we wish to develop a search
strategy that can make use of the entire set
of data taken.

Since the rectangular lattice of $f'$ and $f''$ values has
exactly the same density over the entire orbital plane, the
experimenter could do just as well by searching the data
for Doppler-shifted pulsar behavior at {\bf any} a priori
selected orbital phase domain, including conjunction.  
In this case, small steps must be taken in $a_{p}$, 
but large steps are allowed in $\Phi$.
The only danger is that second order deviation of
$f'$ from the maximum value caused by the now cruder
steps in $\Phi$ (set by the tolerance on $f''$),
could exceed the tolerance set
by the finer steps in $a_{p}$.  With a little calculation,
one can show that the ratio of the step size in $\Phi$
required by $f''$ to that required by the 2nd order
change in $f'$ near maximum, can be given as:
\begin{equation}
{{\delta \Phi_{f''} \over \delta \Phi_{f',2nd}}(\Phi=0,\pi)=
0.00223({650 \over f})^{1/2}({P\over T})^{2} P_{d}^{-1/3}
({2.4M_{\odot} \over M})^{1/6} ({M \over {2.4M_{c}}})^{1/2} (\sin i)^{-1/2}}~.
\end{equation}
Thus, paradoxically, the conjunction searches for millisecond pulsars
are in trouble due to second order changes in $f'$
until the orbital period, $P$, becomes {\it shorter} than
$\sim$21 times the observation time, $T$, unless $f$ is {\it faster} than
650 Hz, or the companion mass is large.   The rationale behind this statement is that the longer
orbital periods allow much {\it larger} values of $\delta\Phi$ at
conjunction before
$f''$ is a problem, i.e., the dependence of $P$ on $\delta f''$
is stronger and dominates the two effects.
Eventually, however, as $P$ becomes much shorter,
the approximation involved in equating
$f'$ and $f''$ to the orbital parameters has
broken down due to the neglect of the higher
derivatives.
 
Before beginning our discussion of short period binaries,
we want to find out how to transition between the two
extremes of searches.  We can do this by focusing our 
attention on the tolerance in 
$\delta \Phi * a_{p}~(\Phi={\pi \over 2},{3\pi \over 2})$ 
given by equation (11).  In the limit of 
$\Omega T \leq \sqrt{42}$, or basically for
all observations of duration {\it less} than one orbital period,
this tolerance is always more stringent than that
given for $\delta a_{p}$ in equation (12).  It is
identical to the tolerance on $\delta a_{p}$ dictated
by the tolerance of $f'$ at the conjunctions ($\Phi = 0,\pi$), when
$f'$ has its extreme values that multiplication by $\delta a_p/a_p$
can lead to (potentially large) fractional variations.  The point
we wish to make here, then, is that {\it no parameter spacing
needs to be finer than this tolerance as the orbital
period, $P$, gets long in comparison to the data stream
length to be analyzed, $T$}.

\subsection{Searches in Short Period Binaries}

Now that the asymptotic value for the
spacing of the orbital parameter space has
been found, we can investigate the
situation for the very short period binary
systems, with approximately one or more
binary orbits occurring during the data
interval, $T$, with the goal of smoothly connecting pulsar 
searches with short and longer orbital periods.
In the situation where one or more orbits occur
during the run time, $T$, a binary orbit causes
a phase modulation of amplitude:
\begin{equation}
Z_{\phi} = {2 \pi f a_{p} \sin i \over c}~~~~(radians)~,
\end{equation}
so that the equation of the data becomes:
\begin{equation}
{g(t) = a \ast noise +
b*(1+\cos (2 \pi ft + \phi_{0} + Z_{\phi} *\cos (\Omega t + \Phi_{0})))}~.
\end{equation}
The Fourier transform of this signal will produce
peaks in the Fourier power
spectrum at $f$, $f\pm {\Omega \over 2 \pi}$,
$f\pm 2{\Omega \over 2 \pi}$, $\cdots$, $f\pm n{\Omega \over 2 \pi}$, $\cdots$,
each with complex Fourier amplitude
proportional to $(\sqrt{-1})^{n}J_{n}(Z_{\phi})$, where the $J_n$
are Bessel functions of n$^{th}$ order, and
a technique has been developed that exploits this
pattern in the Fourier power spectrum (Ransom, Cordes, and
Eikenberry 2003).
In this brute force method, however, the time series is stretched 
to match a particular set of orbital parameters and the mismatch of 
these to the true set generates a particular error vector,
$\delta Z_{\phi}$ in the complex $Z_{\phi}$ plane, constructed, as other 
complex planes, to track the modulus and phase of such vectors.
A residual phase
modulation then ensues because the error vector has
a finite length, and can also migrate during the observation 
time due to orbital period mismatch.  
For the sake of simplicity, we will assume for now that the
phase modulation has a constant amplitude and phase.
Power is lost to this phase modulation
from the main peak, whose height is proportional to:
\begin{equation}
J_{0}^{2}(\delta Z_{\phi}) \approx 1 - (\delta Z_{\phi})^{2} /2~.
\end{equation}
If we are completely ignorant of the orbital phase and
projected orbital radius of
a particular binary system, then we must search over the
$Z_{\phi}$ plane with tile sizes which can be determined
from equation (16), i.e.,
\begin{equation}
\delta Z_{\phi,90\%} = 1/\sqrt{5}~~~~(radians)~.
\end{equation}
Thus for a worst case mismatch of orbital phase modulation of
$1/\sqrt{5}$ radians, the worst case loss of Fourier power
will be only 10\%.

To cover the $Z_{\phi}$ plane efficiently,
we choose a hexagonal tiling with a side
dimension of $1/\sqrt{5}$.  The hexagonal tile
has the same area ($3 \sqrt{3}/10$) as a square tile of 0.721
radians on a side.
For a tile on the plane with a radius vector of
$Z_{\phi} = 2 \pi f a_{p} \sin i/c$, the orbital
phase subtended by this tile can be given
by:
\begin{equation}
\delta \Phi_{Z_{\phi}} \approx {\delta Z_{\phi} \over Z_{\phi}}
= {\sqrt{2\sqrt{3}/5}~c \over 2 \pi f a_{p} \sin i} \approx 
{0.86 c \over 2 \pi f a_{p} \sin i}~~~~(radians)~,
\end{equation}
where the 0.86 factor on the right hand side approximates the 
geometric mean (0.832) between what are, in turn, approximations
for the angle subtended by the hexagonal tile flat-flat and 
diagonal directions 
(to within 3.3\% -- the actual mean phase angle subtended 
by the tiles to the origin of the $Z_{\phi}$ plane {\it is} 
bigger than what the middle of equation (18)
would predict).  We can now compare {\it half} 
the fineness of this stepping in orbital phase to the worst 
case mismatch increments (10\% loss in power) given by equation (11):
\begin{equation}
\alpha \equiv {\delta \Phi (\Phi={\pi \over 2},{3\pi \over 2}) \over
{0.5 \times \delta \Phi_{Z_{\phi}}}} = ({P \over T})^{2}~,
\end{equation}
where the approximation in equation (18) has conveniently set the 
coefficient on the right hand side of equation (19) to unity.
Equation (19) gives us the relaxation factor 
needed for the tile sizes to make the transition from
MSP searches in short, to those in longer period binaries.
We can now treat the entire search with the short period binary
formalism, by letting the tiles expand by a factor
of $\max(\alpha,1)$, as $\alpha$ increases with increasing
orbital period.  It is no accident that 
$2\pi/\sqrt{42} \approx  1 \approx (2\sqrt{3}/5)^{0.5}$, i.e., the 
larger tile size cuts on just when increasing orbital period,
$P$, or decreasing observation time, $T$,
causes the increments in equation (12), 
relative to $a_p$, to exceed those of equation (11).  
 
The step sizes for the orbital period must be small
enough so that the drift in orbital phase caused by
the period mismatch is less than or equal to the phase 
subtended by a tile edge (drawn through the tile center) 
on the $Z_{\phi}$ plane:
\begin{equation}
{\delta ({1 \over P}) T \equiv \delta FT \approx {0.43c \over 4 {\pi}^2 f a_{p} \sin i}~~~~(cycles)~;
{\rm~or:}~~ \delta \Omega \approx {0.43c \over 2 \pi T f a_{p} \sin i}}~~~~(radians~s^{-1})~.
\end{equation}
A point which starts at an offset of $x$, of a tile edge unit, 
along the tile diagonal perpendicular to the (radial) direction to
the origin of
the $Z_{\phi}$ plane, and then {\it moves} by $y$, of a tile 
edge unit, in the same direction due to orbital period mismatch
(approximating the circumferential arc by a straight line)
could easily share/duplicate its Fourier power among {\it three} 
different tiles.  Assuming losses to Fourier power 
add directly (as in a quadrature sum) we can compute the total 
loss of power in the tile of the point's origin, in units of the 
10\% worst case mismatch, as: 
\begin{equation}
{ { \Delta{(P_F)_I} } \over { {(P_F)_I} } } = x^2 + xy + {y^2 \over 3}~.
\end{equation}
The loss of power to the two tiles whose common edge lies
in the path of the point is:
\begin{equation}
{ { \Delta{(P_F)_{II}} } \over { (P_F)_{II} } } = 
3/4 + {(x - 3/2)}^2 + (x - 3/2)y + {y^2 \over 3}~.
\end{equation}
These become equal on the line in the x-y plane given by:
\begin{equation}
y = 2 (1 - x)~, 
\end{equation}
with a $13/12 \times 10\% = 10.8\%$ maximum loss of 
Fourier power occurring for the $y = 1$ (and $x = 1/2$) 
case posited by equation (17).  For points in the
$Z_{\phi}$ plane which begin halfway (radially) between the tile 
center and edge, proceeding parallel (and thus $\sim$circumferentially)
to those already discussed, the total loss of power in the tile of
origin is:
\begin{equation}
{ { \Delta(P_F)_{III} }  \over { (P_F)_{III} } } = 3/16 + x^2 + xy + {y^2 \over 3}~;
\end{equation}
and the loss in the downstream neighboring tile is given by:
\begin{equation}
{ { \Delta(P_F)_{IV}  } \over { (P_F)_{IV} } } = 3/16 + {(x-3/2)}^2 + xy + {y^2 \over 3}~.
\end{equation}
These become equal on the line in the x-y plane given by:
\begin{equation}
y = 3/2 - 2x~,
\end{equation}
with a $5/6 \times 10\% = 8.3\%$ maximum loss of 
Fourier power occurring when $y = 1$ (and $x = 1/4$).

With these steps, we note that the total
amount of computation needed by such a search is 
proportional to the fourth power of $f$ and $T$, two powers 
to cover the $Z_{\phi}$ plane, another power in the step sizes 
for the orbital period, and the last factor due to the size of
each Fourier transform.  Larger worst-case losses will be incurred, of 
course, by ``inter-binning '' the resulting Fourier transforms, 
as much as 19\% (even as in MDK) unless more than two Fourier 
elements are used.  At this point, a more sophisticated inter-binning
is probably worth the extra computation to reduce the loss to 10\% 
(four elements), or even 6\% (six elements).  However, the impact 
of cruder interbinning can be eliminated in two stage post-Fourier
analyses, with candidates overselected in the first stage,
prior to a second stage that accurately finds the frequencies of their
true maximum power, and then uses these as the basis
for a search for related harmonics.  As the loss of
Fourier power produced by the second harmonic ($2f$) structure
in the MSP's pulse profile is {\it quadruple} that for the
fundamental, it is worth including subharmonics in the search
algorithm (see $\S 2.4$).
  
When the boundary between short and long
periods is reached ($P \ge T$), it is understood that
the tile size merely scales by $(P/T)^2$.  If
the same number of cases are done, the area
being covered on the $Z_{\phi}$ plane increases.
Thus, in almost all practical searches, the {\it differential}
number of cases needed for ($P \ge T$) decreases by $(P/T)^2$.

\subsection{Example} 

As a hypothetical test case, we assume that our target pulsar
has a period of 5 ms, i.e., is 4.54 times as fast as the J0737A
spin rate of 44.05 Hz.  In order to calculate the work necessary
to find such a pulsar in a circular orbit with the same period and
(since the inclination of the J0737 system is nearly 90 degrees) 
semimajor axis, we must first calculate the area of the disk on 
the $Z_{\phi}$ plane we need to search.  The radius vector of this 
disk is given by:
\begin{equation}
\rho_m = {{2 \pi f a_p \sin i} \over c} = 1778~(radians),
\end{equation}
where we distinguish $\rho$ as the radius of the disk on
the $Z_{\phi}$ plane, where $Z_{\phi}$ itself is just the
amplitude of the phase modulation as given in equation (14).  
The number of hexagonal
{\it tiles} in each ring can be counted as {\it two} for every $3/\sqrt{5}$
radians (when counting the number of {\it rings} later, we will count
these as only {\it one} ring), with every other tile offset by 1/2 
the flat-flat distance ($\sqrt{3/5} = \sqrt{0.6} \approx 0.775$) in 
the radial direction, and are separated by $3/(2 \sqrt{5})$ radians 
circumferentially:
\begin{equation}
N_t(\rho) = { {2 \pi \rho} \over { 3 / (2\sqrt{5}) } } = 
{ {4 \sqrt{5}} \over 3} \pi \rho =  16,655~({\rho \over \rho_m})~(tiles).
\end{equation}
Dividing $\rho$ by the flat-flat distance, we can count the number of rings:
\begin{equation}
N_r(\rho) =  {\sqrt{15} \over 3}\rho = 2,296~({\rho \over \rho_m})~(rings).
\end{equation}
The total number of tiles in the disk of radius, $\rho$,
can then be estimated by the product of the two, times 1/2,
due to the integral over the linear ramp of the number of 
tiles per ring:
\begin{equation}
N_{disk}(\rho) = {1 \over 2}  N_t(\rho) \times N_r(\rho) =
1/2{{4 \sqrt{5}} \over 3} \pi \rho {{\sqrt{15} } \over 3}\rho = 
{10 \over 9} \sqrt{3} \pi {\rho}^2~. 
\end{equation}
It would have been easier to just take the disk area, $\pi {\rho}^2$,
and divide by the area per tile, $0.3 \sqrt{3}~{\rm{radians}}^2$:
\begin{equation}
N_{disk}(\rho) = {{{\pi {\rho}^2}} \over {0.3 \sqrt{3} } } = 
{10 \over 9} \sqrt{3} \pi {\rho}^2 = 1.91 \times {10}^7({\rho \over \rho_m})^2.
\end{equation}
However, to compute the total amount of processing needed we
need $N_t(\rho)$ to weight by $N_P(\rho)$, the number of trial 
orbital periods, prior to integrating (see just below).

Assuming our observation time, $T$, is 17,669 s, twice the J0737 orbital period
of 8,834.5 s ($P_7$), we will search over a range in orbital frequency, from:
\begin{equation}
\Omega_n = {{2 \pi} \over T} = {\pi \over P_7} = 3.56\times 10^{-4}~(radians~s^{-1}),
\end{equation}
to:
\begin{equation}
\Omega_x = {{2 \pi} \over P_7} = 7.11\times 10^{-4}~(radians~s^{-1}).
\end{equation}
For observation times longer than $2 P_7 = 17,669$ s, the 
relaxation factor from equation (19) applies and the
amount of added computation, relative to that needed for 
$2 P_7$ to $P_7$, is down by a factor of 3. 
As in equations (20) and (27) the spacing of the orbital frequency search 
gets finer as $\rho$ increases, so that the drift of the vector 
on the $Z_{\phi}$ plane does not exceed $1/\sqrt{5}$ radians:
\begin{equation}
\delta \Omega = {1 \over {(\sqrt{5} T \rho)}}.
\end{equation}
When we calculate diminishing trial periods from $T$ to $P_7$,
the number of trial orbital periods needed in a given ring is:
\begin{equation}
N_P(\rho) = {{(\Omega_x - \Omega_n)} \over {\delta \Omega} } = 
2\pi({{1\over P_7} - {1\over 2P_7}})\sqrt{5}T\rho=\sqrt{5} \pi \rho = 
2.5 \times {10}^4({\rho \over \rho_m})~(trial~periods).
\end{equation}
Now we can multiply the number of trial periods per tile, 
by the number of tiles per ring, and finally by the number of rings
to get the total work.  This time we need a factor of 1/3 for
the integration of the ${\rho}^2$ term:
\begin{equation}
N_{FFT}(\rho)  = {1 \over 3} N_P(\rho) N_t(\rho) N_r(\rho) =
{40 \over 27} \sqrt{15} {\pi}^2 {T \over P_{min}}{\rho}^3 = 
3.18 \times {10}^{11}({\rho \over \rho_m})^3.
\end{equation}
Equation (36) shows why it is much easier to detect the slower 
component of double pulsar systems, such as the 0.36 Hz J0737B, 
thus gaining exact knowledge of the orbital period, phase (and
eccentricity), and approximate knowledge of the companion's 
semi-major axis.

\subsection{Mitigating Factors and Other Details}

In the example above the typical size for each FFT would be set by 
an $\sim$2 kHz sampling rate, times 8834 s, or about $N=2^{24}$ 
samples.  
Processor speeds are such that each these can be done in under 
a second.  The problem arises, however, in getting the 300 
billion processor-seconds.  
There is usually enough memory
local to modern processors to accommodate the
relatively short FFT's, so that the computing can proceed
in the ``embarrassingly parallel'' mode.
The large number of trials also requires an extra 25.8 units of 
power to amortize, which is enough to be significant in just one 
trial.  However, the coherent addition of higher and/or 
lower harmonics (e.g., as in M00, equation [1])
can make up for this quickly, particularly for the first three 
harmonics of spin periods of 15 ms or longer.  
Targeted searches which first find the $0.1\%$ to $0.5\%$ 
most significant "seed" events in the Fourier transform, can be 
used to coherently investigate trains of whole, half, and 
1/3 harmonics. These can be evaluated in only one extra 
pass through the Fourier spectrum, with very little loss
by using an interpolation involving at least six complex 
amplitudes.  

Such a technique 
can avoid missing significant periodic signals due
to the statistical fluctuations in the seeds, generally
with an expected mean power level between $10~(-3~+4.5)$ and 
$17~(-4.24~+6)$ (i.e., with otherwise a much greater potential loss 
of power than just 10\%) because down fluctuations in all {\it three} 
seeds are unlikely, and the coherent summing process, once started, is 
insensitive to them.  In addition, the more extensive searches
will, of necessity, have seeds with more power, and hence smaller 
frequency errors, making them more efficient in the harmonic
summation process, particularly when including the whole integer
higher harmonics (and thus whole integer multiples of the
{\it error} in the fundamental frequency).
Still, this amount of computation is prohibitive.

It {\it is} possible that by 
convolving the original Fourier spectrum with perfectly matched 
filters for the phase modulation between one tile and its neighbors
(a process that utilizes short FFT's in memory also local to the
processors [REM02]) that the majority of the longer FFT's do not
have to be computed.  Only six convolution templates need to
be used, each representing a residual phase modulation for 
a vector, with a length of a tile flat-flat ($\sqrt{3}$ tile edges)
on the $Z_{\Phi}$ plane.  
This set of seven tiles forms one ``7tile.''  Next, we can use 
six more templates, for vector offsets of $\sqrt{21}$ tile
edges (e.g., [4.5,$\sqrt{3}/2$] etc.) to go from the original 
tile to the center tiles of the six neighboring 7tiles, and then
we can flesh out these neighboring 7tiles by again
using the six tile diagonal length convolutions, to 
form a ``49tile.''  The steps in orbital period have to as 
fine as needed for the most radially distant tile in the 
49tile.  Since we are assuming that even the large  
FFT's fit entirely within memory local to each 
processor and FFT computation in such circumstances
takes on the order of $log_2({\rm N})$ this technique can only 
reduce the amount of computation required by a factor of four 
at best, and even then only with a much faster,
and likely less effective, harmonic searching algorithm.

\section{Discussion}

Fortunately, there may be reasons to believe that the MSPs
found in DNSBs will, {\it no matter how the MSPs entered
into the pre-DNSB}, have periods longer than 15 milliseconds,
even though shorter period MSP members would not be 
easily detected.  
Although the predominant species of neutron stars (NSs) are
the merger-spawned, weakly magnetized, MSPs with an
initial spin period of $\sim$2 ms (M04a) at first glance
it would seem to be very difficult to incorporate these 
into DNSBs as the MSP companion.  There is not enough 
time for massive stars to capture them before they suffer
their Fe catastrophe SNe, but if this does happen, then
a 2 ms MSP in a DNSB becomes a real possibility.  The 
less massive companions can not evolve into pre merger-target 
white dwarfs, again because the Fe catastrophe of the massive 
star happens too quickly.

However, it {\it is} perfectly feasible for accretion to slow an
NS rotation that is {\it retrograde} wrt the orbital angular 
momentum, a frequent outcome of NS capture as well as post-capture 
merger-induced core collapse, both of which would most likely 
result from binary-binary collisions.  Thus, if an NS {\it is}
captured, then accretion from the pre-Fe catastrophe massive
star could {\it slow} its spin to 50 Hz, or below.  {\it Double} 
merger, and/or single merger and a disruption accretion disk, 
both resulting from binary-binary collisions, could also deliver a 
weakly magnetized MSP {\it and} a post merger companion sufficiently 
massive to progress to an Fe catastrophe SN, thus ``speeding up'' its 
evolutionary clock.  Whether these mechanisms are prevalent enough
to produce the current distribution of DNSBs remains in question.
However, condensed or collapsed core (CC) globular cluster (GC) stellar 
densities don't seem to be required for the Galactic plane to make its 
share of fast ($P \le 15$ ms) i.e., merger-spawned MSPs, and the numbers 
of such MSPs in the two categories are about the same.  However, given 
that the GCs are generally farther away (save for B1913+16 and 
J1811-1736 at 7 and 6 kpc, respectively, closer than 47 Tuc and M4, at 
4 kpc and 2 kpc) and on account of their known dispersion measures, 
their MSPs have been discovered in {\it longer} time series observations.  
This selects {\it against} discovering GC, relative to Galactic plane 
(GP) DNSBs, thus the current GC/GP DNSB ratio of 1 to 7 is not 
unexpected for the same formation mechanism in both locations. 

Some even hold 
that accretion onto a {\it directly} rotating (prograde) NS can also
slow its spin rate \citep{K02}, and there may be some evidence for 
this in the MSP population of the GCs (see below).  In this case,
the DNSB {\it can} gain its MSP through, again most likely, 
binary-binary collisions in which the MSP companion is captured 
(but not likely, except in the case of double merger, {\it formed} by 
white dwarf-white dwarf merger and subsequent core collapse).  
An epoch of accretion ensues which then, relatively quickly, {\it slows} 
the NS spin.  Later, the companion undergoes an Fe catastrophe SN 
and forms the highly magnetized NS.  This opens up the possibility
that DNSBs could have MSP components spinning faster
than $\sim$50 Hz, and the reason none have been found
so far would have to be due to the kinds of selection effects
discussed here.  It is also possible that, because the accretion 
onto the J0737As must {\it always} happen if the progenitors of 
the J0737Bs are to evolve into Fe catastrophe SNe, the 2 ms 
periods may never occur for the J0737As by the time of the
Fe catastrophe SNe of the pre-J0737Bs.

Before getting more specific about the details of the evolution 
of the J0737 binary system, we can still draw a few preliminary 
conclusions.  Observations prohibit the MSP companions from being 
produced {\it directly} by their Fe catastrophe SNe, unless they 
are born so weakly magnetized as to persist as radio live pulsars 
beyond the epochs of their companions' impending Fe catastrophe SNe.
Although some Fe catastrophe SNe have been born
spinning faster than 100 Hz \citep{M98}, and some 3D calculations 
have indicated that a weakly magnetized NS could result 
from one \citep{FW04} the very state of the remnants of 
the last SNe in DNSBs testifies otherwise.  
Out of eight DNSBs, seven have radio-dead companions, and
the last (J0737B) has a $1.2 \times 10^{12}$ G field and is spinning 
so slowly (2.8 s) as to have one foot already in the grave.  In 
addition, the numbers of weakly magnetized Fe catastrophe SNe are 
limited because the gamma ray burst (GRB) rate sets a strict 
{\it lower} limit to the incidence of merger-induced SNe \citep{Sch99}.
(In case there remain any lingering doubts that SN 1987A, 
merger-induced core collapse SNe, and GRBs are intimately 
related, we note, from M04a, that the product of the energy of 
the ``mystery spot,'' $10^{49}$ ergs, and the
beaming factor for GRBs, $10^5$, yields $10^{54}$ ergs,
a fair fraction of the maximum assumed isotropic fluence of
GRBs.) Thus, unless we argue, that in spite of the limited rate, 
some of the seven are similar to Cas A 
\citep{T99}, with no evidence of a magnetic field {it at all}, Fe 
catastrophe SNe can only {\it rarely} produce NS remnants with the
weak fields typical of MSPs, if ever.  Field {\it growth} in pulsars 
due to magnetic/thermal interaction (Blandford, Applegate, \& 
Hernquist 1983) even after hundreds of years, 
must also be considered unlikely, again because of Cas A. 
Thus the eight DNSBs constitute the best evidence so far 
that Fe catastrophe SNe {\it usually} produce remnants with magnetic 
fields well in excess of $10^{10}$ Gauss.

In addition, PSR J0737A has a spin rate of 44 Hz, and, extrapolating 
backward, its initial rate, just prior to the pre-J0737B Fe catastrophe 
SN, was not much higher.  This spin rate is more typical of those 
few MSPs with slower spins (Chen, Middleditch, \& Ruderman 1993, 
hereafter CMR93) and distinctively sharp pulse profiles 
such as M15 A, B, C, and G \citep{A92,CR93} and another 
handful in the other CC GC clusters and Galactic plane, 
that most agree have actually been recycled.  Thus, J0737A must
have been accreting from a star massive enough to undergo
Fe catastrophe core collapse, and in spite of that was still
spinning at only 50 Hz at that epoch some 50 Myr ago.
Taken at face value, this supports the picture of 
predominant, fast (2 ms) merger-induced MSPs, versus the 
slower, recycled (and hence rare) MSPs.  

The ``standard'' model for DNSB formation starts with two massive,
(but not equally so) stars, and requires two different 
types of accretion at two different epochs.  The first involves 
the evolution of the more massive companion star A (the precursor 
to J0737A) and unstable mass transfer onto star B (the precursor of 
J0737B) as a way of avoiding a common envelope inspiral in short 
period binaries \citep{Bv91,Dv04,WK04}.  Star A then undergoes an 
Fe catastrophe SN, leaving J0737A with the helium-rich companion, 
star B, which then evolves and overflows its Roche lobe, beginning 
the second epoch of mass transfer (this time stable) onto J0737A as 
a  high mass X-ray binary system (HMXB).  This era of accretion 
spins up J0737A to an intermediate rotational frequency, while 
reducing its effective magnetic dipole moment.  Star B evolves and 
undergoes an Fe catastrophe SN, leaving the highly magnetized NS, 
J0737B.  Following this logic, {\it if} accretion were the {\it only} 
way to get a rapidly spinning NS, then the DNSBs would constitute 
the best evidence so far that the magnetic field configuration of 
an NS can be modified by accretion.  But, of course, it isn't (M04a).

The assumption implicit in the standard model, is that a longer
interval of accretion in a binary ($\sim 10^9$ yrs) can typically 
spin up an NS to rotate as fast as 500 Hz, 
so that a shorter interval of accretion ($\sim 5 \times 10^7$ yrs) 
while star B evolves toward its Fe catastrophe SN, is sufficient to 
spin it up from 0 to 50 Hz, and reduce its magnetic field to a value 
typical of MSPs.  However, as already mentioned, GRBs require
a certain number of merger-spawned MSPs born with periods near 2 ms, leaving
less and less room in the MSP population for those expected
to be recycled into spinning as fast as 500 Hz.  In addition,
the details of the MSP population in the globulars do not support 
recycling as the origin of the fast MSPs (2-15 ms, M04a).  
So perhaps recycling may so ineffective that it could even have a 
hard time just spinning up J0737A from 0 to 50 Hz in only 
$5 \times 10^7$ yrs.

The similarity of the distribution of MSP spin periods in 
the DNSBs (23 to 104 ms) to that of the ``truly'' recycled MSPs 
in the CC GCs (23 to 111 ms, with the 30.5 ms M15C in common) is 
uncanny, and, if not due to selection effects, could have physical 
meaning.  It may only go so far, however, as J1811-1736 and B1913+16, 
with the longest periods of the eight DNSBs (104 and 59 ms) have the 
broadest (even relative) pulse profiles, W50(ms)/P0(s) = 225 and
207 (ATNF 2004 -- note that W50 is a function of the frequency 
band and time resolution of the observation, so caution must be used),
with the rest scattering 
between 18 and 106.  The same relation for the MSPs M15A-H,
respectively, yields, in (W50/P0, P0) pairs: (18,111), (61,56), (43,31), 
(271,4.8), (172,4.65), (197,4.0), (21,38), and (104,6.7) \citep{A92} 
and thus shows the opposite trend.  The relation in DNSBs 
may be due to varying amounts of accretion, while that in M15 
may be due mostly to age.  However, age may also have affected B1913+16, 
in that it has slowed significantly in just the last $10^8$ years, 
leaving the putative varying spinup-caused direct relation between P0 
and W50/P0 intact only for J1811-1736.  If we ignore J1811-1736, then 
M15C and J0737, with periods of 30.5 and 22.7 ms, have W50/P0s of 43 
and 106, consistent with the trend in the CC GCs, but the evidence is 
sparse, {\it and} M15C belongs to both distributions, so the issue 
of similarity in the trend of W50/P0 between the DNS, and the CC GC MSPs
remains undecided.  

The W50/P0 versus P0 relationship for the {\it entire} pulsar population has a 
power law trend with an index near -0.8, indicating wider pulse profiles with 
decreasing period, until it reaches the (mostly recycled) MSPs at P0 = 0.1 s, 
at which point there are many pulsars with pulse profiles that
are too narrow, such as M15A, C, and G. These start a new trend with roughly 
the same power law, that connects to the MSPs with periods shorter than
23 ms.  There is a gap here, but this is due to not having W50s for the 
28.8 ms pulsar, J0621+1002, and the 23.1 ms pulsar, J1804-0735 in the GC NGC 
6539, and, of course, J0737A will also help fill it.  The new trend is about a 
factor of ten below (toward narrower pulse profiles) any W50/P0 value of 
the extrapolation from the longest period pulsars, which is good because 
the continuation of the upper track exceeds 1000 near P0 = 10 ms, thus becoming 
unphysical (the $\sim$1.6 ms pulsars B1937+21 and B1957+20 fall below even the 
{\it lower} track).  The dichotomy of the slow/narrow versus the 
fast/broad pulsars in M15 and the other CC GCs then, can also be viewed 
as consequence of this trend, and the near order of magnitude geometric
mean difference factor between the sets of long and short spin periods.  
So instead of asking why some pulse profiles are wider than others, 
it may make more sense to ask why the distribution of spin {\it periods} 
of the CC GCs is bimodal.  

All of this begs the question of if and how the recycling history
of seven of eight DNSB MSPs differs from that of the CC GC MSPs,
and how is the configuration of the NS magnetic field affected 
by each and why?  The recycled GC MSPs, including those in M15, 
certainly had no several-solar mass helium star from which to 
accrete, so they must have managed with leaner
pickings, or, again, the accretion itself slowed them.  
What is apparent from ``normal'' accretion is that, 
the more magnetized the NS, the slower the equilibrium 
spin frequency, from MSPs (though this evidence is sparse), 
through the HMXBs, and on down to AXPs/magnetars.  
Somehow, the field manages to get modified and we're still 
unsure if we have ever witnessed this process.  Thus, if only 
{\it non-braking} accretion occurs, there obviously have to be a few 
different states, from ``super-critical'' on down, that have different 
effects on NS magnetic field configuration, and it is 
clear that particular extreme is not happening in HMXBs 
such as Cen X-3 and SMC X-1, or even Her X-1.

Both the standout $\sim$1.6 ms pulsars, B1937+21 \citep{BKHDG} 
and B1957+20 (Fruchter, Stinebring, \& Taylor 1988), have
very narrow (double) pulse profiles consistent with magnetic to 
(single) rotational pole migration due to accretion stresses during 
recycling, without their (now very close together) polar surface field 
{\it strengths} being significantly attenuated (CR93).  As mentioned 
above, these are both located at the bottom extreme of the high frequency
end of of the lower W50/P0 vs P0 track.  {\it If} accretion {\it can} 
slow the spin of most NSs, then it is tempting to say that {\it these two} 
quite possibly form the {\it entire} class of (initially) strongly 
magnetized, and later, recycled MSPs, with {\it all} of the rest having 
weak magnetic fields at birth, through recycling (or not), and afterward.  
In any case, they are the rarest of rare pulsars, being first selected 
against, by as much as a factor of a hundred, because they were offspring 
of Fe catastrophe SNe (M04a) and then (perhaps) again by another 
factor of 10, because they somehow had to have very special accretion/field 
configuration-modifying circumstances.  
The number of ``standard'' MSPs exceeds that of the ``non-standard'' 
MSPs, B1937+21 and B1957+20, by about the same factor that 
the number of merger-induced SNe exceeds the number of Fe 
catastrophe-induced SNe, lending support to the possibility that 
accretion can act differently on the fewer strongly magnetized 
NSs than it does on their much more numerous, but
{\it less} magnetized cousins.
Still we do not yet know {\it how} accretion spins up a few 
strongly magnetized NSs, such as B1937+21 and B1957+20, to very 
high rotation rates, and why it is {\it not} all that effective at doing 
the same with the weakly magnetized \citep{A82}, but pulsar recycling 
appears to work this way.  Studies of this problem are still in their
infancy \citep{C04}.  For now, however, we must conclude that there 
is a remote possibility that a 1.6 ms pulsar will be a member of a 
DNSB.

The last remaining holdout is J0034-0534, 
a pulsar in a 1.6 day binary with a low mass companion in a 
circular orbit of $\sim$10 lt-s \citep{B94}.  This unique
pulsar has a record low magnetic field and near record 
broad pulse profile,
a 1.87 ms spin period, and a very steep radio spectrum 
\citep{AT04}, all of which make it a candidate for a merger of 
a white dwarf with a weakly magnetized NS \citep{M04b}.  
The spin period is shorter (just) than the 2.1 ms 
minimum period for merger-spawned pulsars \citep{Ca00} 
and longer than $\sim$1.6 ms (the ``Chen Gap,'' 
after Kaiyou Chen).  The pulse profile, however, is not
dramatically wider in W50/P0 than the mean of the low P0 distribution 
extrapolated to this frequency\footnote{Ignoring the erroneous values of
998 for J2051-0827 (the true value is about 60 -- Stappers et al.~1996) 
and the additionly {\it unreferenced} value of $\sim$3600 for J1807-2459!}
but, at this point there is very little {\it room} for increase.  
Again, there is only a very remote possibility that such a pulsar will 
be a member of a DNSB.

The more likely cousins to J0737, and certainly the easiest to find,
short of detecting the younger, slower, strongly magnetized member,
may be other similar systems, with a modestly fast MSP, like J0737A, 
and shorter or longer orbital periods, but at this point, we can not 
rule out selection effects favoring slower MSPs over faster.
Nevertheless, we take, as an example, a pulsar with the same total 
mass of 2.58 solar, in a binary system with an orbital period of 
1472 s, six times shorter than $P_7$, the 8,834 s orbital period of 
J0737, {\it and} our assumed observation interval.  To keep the argument 
simple, we relax the constraint on 
total mass for orbital periods longer than 1,472 s.  Otherwise, $\rho_m$
scales as $(P/(1,472 s))^{2/3}$ and the work scales as 
$(P/(1,472 s))^2$ as $P$ gets larger.  For this case, 
$a_p \sin i = 0.43$ lt-s and $\rho_m = 119$
radians.  The amount of computation needed for orbital periods 
longer than $P_7$, relative to that required for $P_7$ 
to $P_7/6$, is down by a factor of 15.
The number of trial orbital periods needed at $\rho_m$ for the
range of 8,834 s to 1472 s is $N_P = 8,332$.  The
number of tiles/ring at $\rho_m$, $N_t = 1,111$.  And finally,
the number of rings at $\rho_m$, $N_r = 153$.  
In this case the amount of computation needed 
at $\rho_m$ (recall the factor of 1/3 in equation [36])
is only $4.73 \times 10^8$, nearly a factor of
a thousand less, and very accessible to modern 
computers.  

The reduction is due to two factors,
the first being the decrease by a factor of 3.3 of the
semimajor axis, which produces a factor of 36 when
cubed. The second is the factor of 
4.54 reduction in pulsar frequency, from 200 to 44 Hz, 
producing a factor of 94.  Together these produce a factor 
of 3,375, or the cube root (15) in $\rho_m$ 
({\it unrelated} to the factor of 15 which applies
to the long-to-infinite orbital period cases posited above).  
Against this, we used more trial orbital periods by a factor of
5 in the 1,472 s example, so our net reduction 
in calculation is about 675.
The Fourier transforms are also shorter for this example, but
the bad news is that the associated harmonic searches will work well
only for 22 Hz and lower spin frequencies (for the 2nd harmonic), and 
14 Hz and slower for the third harmonic.
The good news is that the work required for
fixed mass scales as $P^2$ as $P$ gets smaller, while, at the 
same time, the pulsar orbit is more likely to be circular.
Other alternatives include just reducing the observation time
analyzed by a factor of 4, to 2,208 s, which would sacrifice
a factor of 2 in signal-to-noise ratio, but which would also
change the fully computationally intensive upper limit period
from 8,834 s down to 2,208 s, reducing the amount of computation 
to 2,208 s by a nearly a factor near 2.3, in addition to the 
factor of 256 reduction everywhere, counting FFT lengths.  
Like the original example in $\S 2.3$ such a search would be 
effective up to the third harmonic of a 15 ms pulsar.

Of course, if we want to search for MSPs with black
hole companions, the total mass of the binary system will
almost certainly exceed three solar, with the required 
amount of computation rising steeply with black hole
mass.

\section{Conclusion}

In this work we have derived the form of the transition between
acceleration binary pulsar searches and truly short period
binary pulsar searches.  We have also shown here that although 
the amount of computation needed to search for MSPs in short 
period-high companion mass binary systems appears to be 
stupendous, rising as the fourth power of both the 
observation time and targeted upper limit pulse frequency, 
many such searches are now within our grasp, given intelligent 
choices of ranges for the unknown parameters.
Recent advances in the understanding of the nature of supernovae
and their compact remnants (e.g., M00, M04a, and this work, $\S 3$) 
can play a crucial role in narrowing otherwise prohibitively expensive 
computer searches for MSPs in binary systems.  The easiest, and
perhaps, most likely candidates for more DNSBs are similar to 
J0737 (though this is far from certain) with a weakly magnetized 
NS spinning only moderately fast. The possibility of finding any 
pulsar in a DNSB, with a $\sim$10 to 1.5 ms period, appears 
to be very remote, but many crucial details of DNSB formation and
the pulsar recycling process remain mysteries.

The astonishing similarity between the MSP periods in the DNSBs
and the those of the longer period MSPs found only in 
the CC GCs, including the one MSP/DNSB in common, M15C, calls
the standard model, which requires back accretion from
a several solar mass, pre-Fe catastrophe SN helium star,
into question, in spite of what Lorimer et al.~(2004)
would have us believe.  Other options include MSP 
formation via binary-binary collisions and spin{\it down} 
via the reverse accretion.

For the short period X-ray binary systems in particular, the
range of parameters for such searches can be narrowed with
exact knowledge of the orbital period and approximate 
knowledge of the pulsar candidate's projected semi-major axis
and orbital phase.  Searches for MSPs in binary
systems where one pulsar is already known, can be
narrowed even further, as the orbital period, phase, and 
eccentricity are known exactly.  However, slowly-rotating
(non-MSP) pulsar companions in DNSBs do not require a 
very sophisticated search in order to set stringent upper limits, and
eventually Relativity will likely provide {\it all} the
orbital elements for the undetected companion NS,
although these have much shorter radio lifetimes.
Finally, even though this work does not discuss
how to treat eccentricity, it can serve as a guide in
the cases for which it is known exactly.

\acknowledgements

We thank Scott Ransom for discussions. This work was performed 
under the auspices of the Department of Energy,
and supported in part by the Institute for Nuclear and Particle 
Astrophysics and Cosmology (INPAC).

\end{document}